\documentclass[sigconf, nonacm]{acmart}
\usepackage{xspace}
\usepackage{booktabs}
\usepackage{tabularx}
\usepackage{multirow}
\usepackage{array}
\usepackage{enumitem}  % 用于自定义列表样式
\usepackage{titlesec}  % 控制章节标题样式
\usepackage{hyperref}  % 处理超链接（如果有）
\usepackage{algorithm}
\usepackage{algpseudocode}
\usepackage{amsmath}
\usepackage{mathtools}
\usepackage{listings}
\usepackage{xcolor}
\usepackage{textcomp}

\definecolor{dkgreen}{rgb}{0,0.6,0}
\definecolor{gray}{rgb}{0.5,0.5,0.5}
\definecolor{mauve}{rgb}{0.58,0,0.82}

\DeclareMathOperator*{\argmin}{argmin}

\newcommand\vldbdoi{XX.XX/XXX.XX}
\newcommand\vldbpages{XXX-XXX}
\newcommand\vldbvolume{14}
\newcommand\vldbissue{1}
\newcommand\vldbyear{2020}
\newcommand\vldbauthors{\authors}
\newcommand\vldbtitle{\shorttitle} 
\newcommand\vldbavailabilityurl{https://github.com/wotchin/AnDB/tree/with-semantic}
\newcommand\vldbpagestyle{plain} 

\newcommand{\hi}[1]{\vspace{.25em}\noindent{\bf {#1}}\xspace}

\begin{document}
\title{AnDB: Breaking Boundaries with an AI-Native Database for Universal Semantic Analysis}

%%
%% The "author" command and its associated commands are used to define the authors and their affiliations.
\author{Tianqing Wang}
\affiliation{%
  \institution{Huawei}
  \city{Toronto}
  \state{Canada}
}
\email{wangtianqing2@huawei.com}

\author{Xun Xue}
\affiliation{%
  \institution{Huawei}
  \city{Toronto}
  \state{Canada}
}
\email{xun.xue@huawei.com}

\author{Guoliang Li}
\affiliation{%
  \institution{Tsinghua University}
  \city{Beijing}
  \state{China}
}
\email{liguoliang@tsinghua.edu.cn}

\author{Yong Wang}
\affiliation{%
  \institution{Huawei}
  \city{Hangzhou}
  \state{China}
}
\email{wangyong308@huawei.com}

% \author{Alice}
% \orcid{0000-0002-1825-0097}
% \affiliation{%
%   \country{Iceland}

% \author{Bob}
% \orcid{0000-0001-5109-3700}
% \affiliation{%
% \email{vb@rocquencourt.com}

\begin{abstract}
In this demonstration, we present AnDB, an AI-native database that supports traditional OLTP workloads and innovative AI-driven tasks, enabling unified semantic analysis across structured and unstructured data. While structured data analytics is mature, challenges remain in bridging the semantic gap between user queries and unstructured data. AnDB addresses these issues by leveraging cutting-edge AI-native technologies, allowing users to perform semantic queries using intuitive SQL-like statements without requiring AI expertise. This approach eliminates the ambiguity of traditional text-to-SQL systems and provides a seamless end-to-end optimization for analyzing all data types. AnDB automates query processing by generating multiple execution plans and selecting the optimal one through its optimizer, which balances accuracy, execution time, and financial cost based on user policies and internal optimizing mechanisms. AnDB future-proofs data management infrastructure, empowering users to effectively and efficiently harness the full potential of all kinds of data without starting from scratch.
\end{abstract}

\maketitle

\pagestyle{\vldbpagestyle}
\begingroup\small\noindent\raggedright\textbf{PVLDB Reference Format:}\\
\vldbauthors. \vldbtitle. PVLDB, \vldbvolume(\vldbissue): \vldbpages, \vldbyear.\\
\href{https://doi.org/\vldbdoi}{doi:\vldbdoi}
\endgroup
\begingroup
\renewcommand\thefootnote{}\footnote{\noindent
This work is licensed under the Creative Commons BY-NC-ND 4.0 International License. Visit \url{https://creativecommons.org/licenses/by-nc-nd/4.0/} to view a copy of this license. For any use beyond those covered by this license, obtain permission by emailing \href{mailto:info@vldb.org}{info@vldb.org}. Copyright is held by the owner/author(s). Publication rights licensed to the VLDB Endowment. \\
\raggedright Proceedings of the VLDB Endowment, Vol. \vldbvolume, No. \vldbissue\ %
ISSN 2150-8097. \\
\href{https://doi.org/\vldbdoi}{doi:\vldbdoi} \\
}\addtocounter{footnote}{-1}\endgroup

\ifdefempty{\vldbavailabilityurl}{}{
\vspace{.3cm}
\begingroup\small\noindent\raggedright\textbf{PVLDB Artifact Availability:}\\
The source code, data, and/or other artifacts have been made available at \url{\vldbavailabilityurl}.
\endgroup
}

\section{INTRODUCTION}

In recent years, the data landscape has evolved with the emergence of data lakehouses, a hybrid architecture combining the scalability of data lakes with the performance and management features of data warehouses ~\cite{Zaharia2021LakehouseAN}. However, despite their strengths in data storage, data lakehouses struggle to effectively analyze unstructured data, which constitutes 90\% of global data ~\cite{IDC_report} and remains a largely untapped resource ~\cite{BOXblog90}. Unstructured data, encompassing text, images, audio, and video, presents significant challenges due to its complexity and heterogeneity, often leading to data silos, inefficiencies, and missed opportunities for actionable insights ~\cite{Madden2024DatabasesUQ}.

The rapid advancement of Artificial Intelligence (AI), particularly Large Language Models (LLMs), offers promising solutions to bridge the gap between semantic queries and unstructured data. Techniques like Retrieval-Augmented Generation (RAG) enhance LLMs by integrating external knowledge, reducing hallucinations, and improving accuracy ~\cite{rag2020}. However, RAG systems are limited by their focus on retrieval, filtering, and ranking, with inadequate support for advanced operations like aggregation and join. %Additionally, natural language interfaces often suffer from ambiguities, as users may not provide precise queries, and the "select-where-sort" RAG formulations fail to optimize natural language inputs fully.

Recent research has sought to integrate traditional database methodologies with AI technologies. For instance, the "LOTUS" system ~\cite{patel2024semanticoperators} introduces semantic operators (e.g., semantic join, semantic group) to enhance AI-driven analysis, while "PALIMPSEST" ~\cite{LiuPalimpzestOA} enables LLM-powered queries through a declarative process. Notwithstanding these advancements, significant challenges remain in the precise interpretation of user requirements, and the implementation of effective feedback mechanisms for system calibration. A case in point is the "PALIMPSEST" system, which necessitates user programming intervention, demonstrates inadequate coverage in its execution plan generation, and disregards conventional optimization strategies for data scan operations, etc.

In this demonstration, AnDB addresses these limitations by providing an intuitive SQL-like interface through its \texttt{SQL Engine}, enabling seamless analysis of structured and unstructured data. The \texttt{Query Optimization} component generates multiple execution plans, with an optimizer selecting the most efficient one. Results are delivered relationally via the \texttt{Execution Engine}, ensuring clarity regardless of data structure. Additionally, the \texttt{Calibration} component estimates result accuracy and core metrics (e.g., execution time, token usage), feeding calibration information back into the optimizer to improve future performance.

\section{System Overview}
\begin{figure}[htbp]
    \centering
    \includegraphics[width=0.3\textwidth]{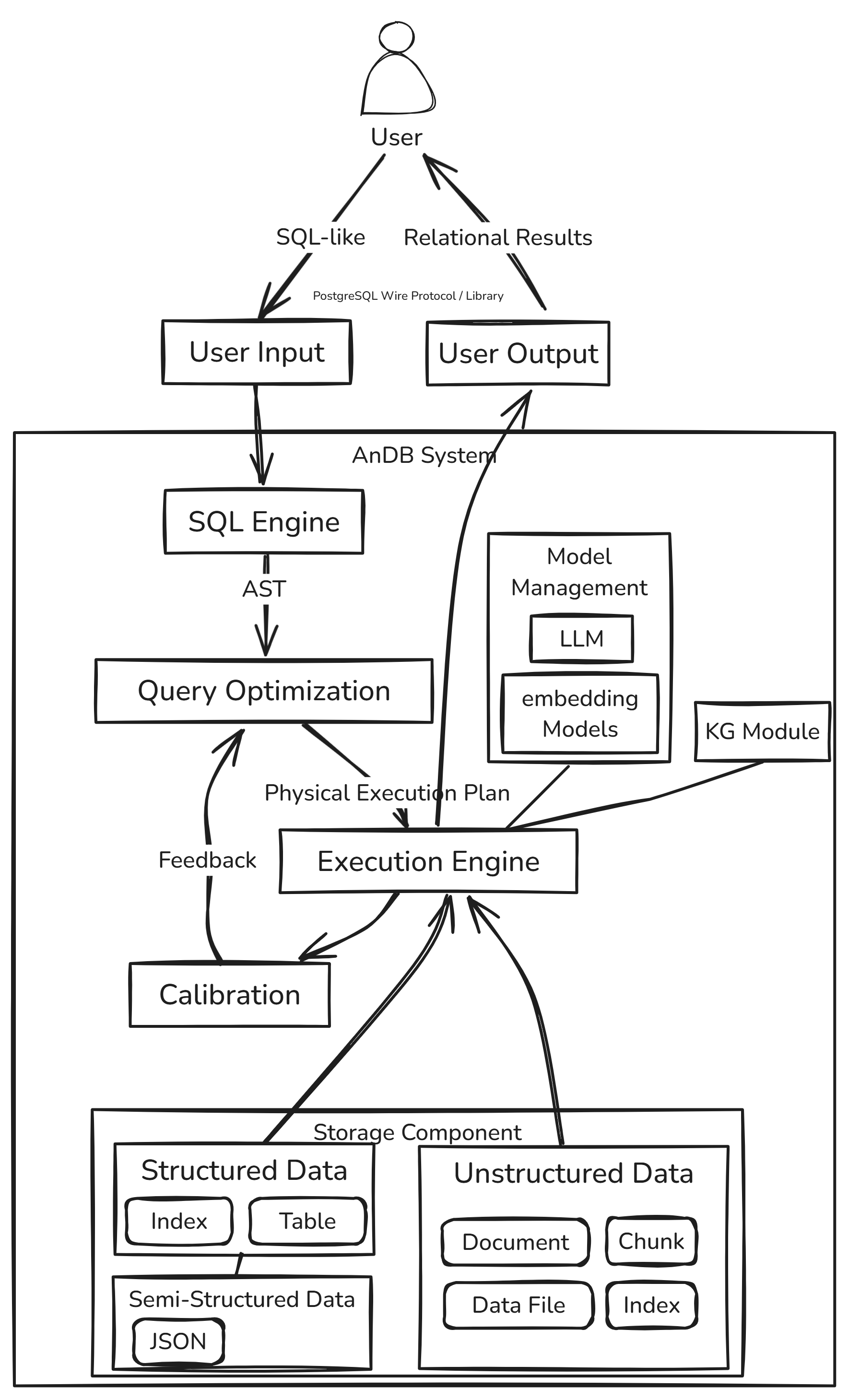}
    \caption{AnDB Architecture Overview: the comprehensive workflow of the AnDB system}
    \label{fig:AnDB_Architecture}
\end{figure}

As illustrated in \autoref{fig:AnDB_Architecture}, AnDB comprises several core components, including the \texttt{SQL engine}, \texttt{query optimization engine}, \texttt{execution engine}, \texttt{storage component}, and auxiliary modules such as the \texttt{model management} and \texttt{Knowledge Graph (KG) module}. Users can seamlessly interact with AnDB using SQL-like statements for unstructured data analysis and standard SQL for relational data queries.

% Thanks to its database-centric architecture, SQL-like interface, and relational result sets, AnDB is fully compatible with the PostgreSQL wire protocol. This compatibility allows users to use any PostgreSQL client to interact with AnDB, significantly reducing the cost of adoption and the learning curve.

% To put it into a user perspective, AnDB operates as a black-box solution, eliminating the need to worry about low-level details such as unstructured data storage, precise query language input, data governance workflow optimization, client software installation, or integration with existing software architectures. In essence, AnDB provides a comprehensive, all-in-one solution—everything you need for modern data management and analysis.

\subsection{SQL Engine}
\hi{SQL Grammar Tokens.} In addition to standard SQL statement tokens, AnDB introduces additional tokens to express user requirements for semantic analysis tasks accurately. In \autoref{tab:tokens}, these tokens are categorized into Semantic Tokens and Auxiliary Tokens, enabling end-to-end semantic analysis with enhanced flexibility and precision.

% \setlength{\tabcolsep}{4pt} % default 6pt
% \begin{table}[t]
%   \caption{Additional SQL Grammar Tokens.}
%   \label{tab:tokens}
%   \centering
%   {\small
%   \begin{tabular}{>{\ttfamily}l>{\ttfamily}l>{\raggedright\arraybackslash}p{5cm}}
%     \toprule
%     Token & Category & Description \\
%     \midrule
%     \verb|PROMPT| & semantic & Specifies semantic requirements based on the user's prompt text. Applicable in SELECT, WHERE, FROM clauses. \\
%     \verb|SEM_MATCH| & semantic & Incorporates semantic matching logic and conditional parameters, facilitating advanced querying capabilities beyond traditional exact matches. Applicable in JOIN\_ON, WHERE, HAVING clauses. \\
%     \verb|SEM_GROUP| & semantic & Prompts for Semantic Aggregation. Applicable in SELECT, GROUP\_BY. \\
%     \verb|TABULAR| & auxiliary & Enables users to define a custom structured schema for unstructured data, bridging the gap between schema-less data and structured querying. Applicable in FROM clause. \\
%     \verb|FILE| & auxiliary & Specifies a path of the unstructured file for processing. Applicable in FROM clause. \\
%     \verb|DIRECTORY| & auxiliary & Specifies a directory containing multiple unstructured files, treating them as a unified dataset for batch processing. Applicable in FROM clause. \\
%     \bottomrule
%   \end{tabular}
%   }
% \end{table}

\setlength{\tabcolsep}{4pt} % default 6pt
\begin{table}[t]
  \caption{Additional SQL Grammar Tokens.}
  \label{tab:tokens}
  \centering
  {\small
  \begin{tabular}{>{\ttfamily}l>{\ttfamily}l>{\raggedright\arraybackslash}p{4.5cm}}
    \toprule
    Token & Category & Description \\
    \midrule
    \verb|PROMPT| & semantic & Specifies semantic requirements based on user prompts. Applicable in SELECT, WHERE. \\
    \verb|SEM_MATCH| & semantic & Enables semantic and approximate matching with conditional parameters. Applicable in JOIN\_ON, WHERE, HAVING. \\
    \verb|SEM_GROUP| & semantic & Prompts for semantic aggregation. Applicable in SELECT, GROUP\_BY. \\
    \verb|TABULAR| & auxiliary & Defines a custom schema for unstructured data. Applicable in FROM. \\
    \verb|FILE| & auxiliary & Specifies a file name for unstructured data. Applicable in FROM. \\
    \verb|DIRECTORY| & auxiliary & Specifies a directory name for batch processing of unstructured files. Applicable in FROM. \\
    \bottomrule
  \end{tabular}
  }
\end{table}

\hi{Logical Plan.} AnDB processes relational data in strict adherence to traditional relational concepts, ensuring compatibility with existing systems. Semantic tokens in AnDB are internally translated into logical plans, naturally bridging the gap between unstructured data query semantics and standard database operators. To achieve this, AnDB introduces several novel operators, e.g., \textbf{Transform}, which resembles a map operation but supports flexible schema based on user prompts. Unlike traditional map operations in MapReduce, which preserve dimensionality, the \textbf{Transform} operator can expand or contract relation dimensions, necessitating a new conceptual framework. 

Particularly, systems that rely exclusively on isolated prompt fragments while failing to incorporate global contextual information consistently exhibit coherence deficiencies. In contrast, AnDB addresses this limitation by comprehensively considering all prompt text parameters (e.g., PROMPT, SEM\_GROUP, and SEM\_MATCH), better understanding user requirements.
% In database systems, a logical plan outlines the high-level execution strategy without specifying implementation details, such as sorting algorithms, index usage, or scan methods. In contrast, a physical plan incorporates these granular details, reflecting decisions made during query optimization.In database systems, a logical plan outlines the high-level execution strategy without specifying implementation details, such as sorting algorithms, index usage, or scan methods. In contrast, a physical plan incorporates these granular details, reflecting decisions made during query optimization.

\hi{Physical Plan.} AnDB integrates prompt requirements into conventional relational operators, leveraging their inherent flexibility. However, a key challenge lies in designing efficient and effective physical operators to implement logical plans. For example, a logical join operator can be implemented through multiple physical operators, such as nested-loop-join, hash-join, and sorted-merge join. As semantically equivalent execution paths, providing diverse physical operator implementations is essential to support the query optimizer in selecting an optimal execution plan.

\autoref{tab:logical_physical_operators} highlights some key physical operators corresponding to their logical operators, demonstrating how AnDB tackles the issue while maintaining flexibility and performance.

\begin{table*}[htbp]
  \centering
  \caption{Logical Operators and Corresponding Physical Operators}
  \label{tab:logical_physical_operators}
  \renewcommand{\arraystretch}{1.2}
  \small % 缩小字体
  \begin{tabularx}{\textwidth}{>{\centering\arraybackslash}m{2cm} >{\centering\arraybackslash}m{3cm} X}
    \toprule
    Logical Operator & Physical Operator & Description \\
    \midrule
    \multirow{4}{*}{Scan}       
    & SemanticScan & Scans unstructured data and extracts it into a tabular format based on SQL context and semantics. \\
    & FileScan & Reads raw bytes of unstructured data directly from storage. \\
    & DirectoryScan & Reads raw bytes of unstructured data from a directory as a unified dataset. \\
    & EmbeddingScan & Scans the child operator's data and embeds it into a vector using an embedding model. \\
    \midrule
    \multirow{2}{*}{Transform} 
    & SemanticTransform & Transforms the child operator's data using an LLM and user-provided or built-in prompts. \\
    & CodeExecution & Generates executable code via LLM based on the operator's target, runs it on child data, and passes results to the parent operator. \\
    \midrule
    \multirow{4}{*}{Join} 
    & NestedLoopJoin (vector similarity) & Performs nested loop join using vector similarity for value comparisons. \\
    & HashJoin (vector similarity) & Performs hash join using vector similarity for value comparisons. \\
    & SortedMergeJoin (vector similarity) & Performs sorted-merge join using vector similarity for value comparisons. \\
    & SemanticJoin & Compares two columns' values by interacting with an LLM. \\
    \midrule
    Group & SemanticCluster & Groups data using unsupervised clustering algorithms. \\
    \midrule
    Filter & SemanticFilter & Filters out irrelevant data by interacting with an LLM. \\
    \bottomrule
  \end{tabularx}
\end{table*}

In addition to the direct one-to-one mappings presented in \autoref{tab:logical_physical_operators}, AnDB also supports the implementation of logical operators through a sequence of physical operators, offering enhanced flexibility and optimization potential. For instance, in the case of the join operation, the process begins with applying \textbf{SemanticTransform} to reformat and normalize the two tables into a unified schema. Subsequently, traditional join methods, such as the nested-loop-join, are applied for the tabular tuples, mirroring the approach used in conventional databases.

\subsection{Query Optimization}

%Many semantic query engine, like ""Regarding LLM, there is an intrinsic connection between token usage and computational cost. However, AnDB argues that makes a simple question more complicated. Hence, AnDB introduces a novel approach for balancing financial cost, accuracy, runtime, user policy, and calibration by a unified cost model. Since financial cost is strongly correlated with execution time, which can be regarded as an intermediate factor, AnDB proposes a cost model to formalize and simplify the trade-off among all considered factors, 

Although several innovations, like "PALIMPSEST", propose balancing financial cost, accuracy, and runtime, we contend that runtime is merely an intermediate factor, as the financial cost strongly correlates with end-to-end execution time. Therefore, we adopt a "less-is-more" strategy to redefine the problem. Notably, we introduce a normalized cost model that defines the trade-off between these factors, which is interpretable and natural to integrate with the existing relational database optimizer. Specifically, we define the accuracy-cost matrix and token vector as follows.

Let \( \mathbf{M} \) represent the accuracy-cost matrix, where the elements correspond to the coefficients for accuracy (\( A \)) and cost (\( C \)) for both input and output tokens:

\[
\mathbf{M} = \begin{bmatrix}
A_{\text{input}} & A_{\text{output}} \\
C_{\text{input}} & C_{\text{output}}
\end{bmatrix}
\]

We also define the token vector \( \mathbf{t} \) as the number of input and output tokens processed by a semantic operator \( o \):

\[
\mathbf{t} = [ t_{\text{input}}, t_{\text{output}}]
\]

Given that semantic operators process \( N \) data tuples, the cost-accuracy trade-off for a semantic operator \( o \) is formalized as:

\begin{equation}
\begin{split}
\Gamma(o) &= \underbrace{C_{\text{input}} \cdot t_{\text{input}} + C_{\text{output}} \cdot t_{\text{output}}}_{\text{Cost Term}} \\
&\quad + \lambda \cdot \underbrace{\left(\frac{1}{A_{\text{input}} \cdot t_{\text{input}} + A_{\text{output}} \cdot t_{\text{output}}}\right)}_{\text{Accuracy Loss Term}}
\end{split}
\end{equation}

where \( \lambda \in [0, +\infty) \) is a tuning parameter that controls the trade-off between cost and accuracy. A higher value of \( \lambda \) prioritizes accuracy, while a lower value emphasizes cost reduction.

The core idea of this formulation is to explicitly minimize the weighted total cost while introducing accuracy as a loss term into the optimization objective, which penalizes low accuracy resulting from cost reduction. This design aligns more closely with the traditional database optimization logic. Hence, the optimization of a hybrid execution plan, \( P \), combining both semantic and traditional operators, where \( \chi(o) \) is the execution cost of a traditional operator \( o \), is then formulated as:

\begin{equation}
\argmin_{P \in \mathcal{P}} \left( \sum_{o \in P_{\text{semantic}}} \Gamma(o) + \sum_{o \in P_{\text{traditional}}} \chi(o) \right)
%\argmin_{P \in \mathcal{P}} \left( \sum_{o \in P_{\text{semantic}}} \Gamma(o) + \sum_{o \in P_{\text{traditional}}} \Chi(o) \right)
%\argmin_{P \in \mathcal{P}} \sum_{o \in P} \Gamma(o) + \text{TraditionalOperatorCost}(P)
\end{equation}

By using cost models, AnDB can select an execution plan that minimizes financial costs while achieving the desired level of accuracy. % This offers a more efficient and cost-effective approach to database query optimization.

\subsection{Execution Engine}
LLM invocations primarily rely on API integration. However, if local deployment is required, there is a lot of significant potential for optimization in areas such as resource scheduling, GPU utilization, and in-memory data caching. Additionally, AnDB has a lightweight knowledge graph module, inspired by GraphRAG ~\cite{edge2024localglobalgraphrag}, to aid in reasoning, as graphs effectively represent entity relationships. The execution engine can retrieve structured, semi-structured, and unstructured data from disk. A key optimization involves indexing unstructured files, such as extracting keywords from images using multimodal models (e.g., FUYU-8b~\cite{li2023otterhdhighresolutionmultimodalitymodel}) and preprocessing documents (e.g., chunking) before storage because performing these tasks during execution would be inefficient.

\subsection{Calibration}
The calibration component parallels the reflection and self-critique aspects of the agentic AI framework, which integrates two approaches: (1) \underline{User-based:} This approach utilizes accuracy feedback provided by the user following query execution. 
% It is both simple and effective, as it directly captures the user's preferences. However, it lacks automation, as it relies on manual intervention, limiting its intelligence for the user. 
(2) \underline{Model-based:} This method employs cross-validation across multiple LLMs and built-in rules, offering a higher level of automation. 
% However, it incurs greater costs, and as a result, it is used sparingly, primarily for spot checks rather than being applied to all queries. Instead, AnDB employs sampling to determine when to apply this method.

\section{DEMONSTRATION OVERVIEW} 
We showcase AnDB's functionalities for processing unstructured data.

\hi{Dataset}. Two unstructured text documents, neurips\_2023.txt, and neurips\_2024.txt, were scraped from papers published at NeurIPS conferences in 2023 and 2024.

\hi{Scenario 1}: Classification and Aggregation. The following three statements are nearly of semantic equivalence but differ in the execution plans. \underline{\it Statement a} represents a simple retrieval scenario, similar to what is typically processed by RAG systems. \underline{\it Statement b} demonstrates a semantic retrieval scenario that does not provide a specific schema. AnDB is required to implicitly infer the schema and perform retrieval based on global semantics. \underline{\it Statement c} is a scenario where the user specifies a schema. AnDB only needs to transform unstructured data into structured data according to the user's requirements and then apply traditional data operators for aggregation. %Their execution plans and workflows are illustrated in \autoref{fig:scenario1}.

\begin{lstlisting}
-- Statement a: simple query scenario
SELECT PROMPT("Analyze technical areas and count the number of publications in each area.") 
FROM FILE("neurips_2024.txt");

-- Statement b: implicit schema inference
SELECT PROMPT("Analyze technical areas"), count(1)
FROM FILE("neurips_2024.txt")  -- no schema
GROUP BY PROMPT("Count the numbers of publications in each area");

-- Statement c: given a specific schema
SELECT count(area), SEM_GROUP(title, "Area of publications", 5) /* 5 is optional, which means divided into five groups */ as area
FROM TABULAR(PROMPT("title of the paper") as title FROM FILE("neurips_2024.txt"))
GROUP BY area;
\end{lstlisting}

% \begin{figure}[htbp]
%     \centering
%     \includegraphics[width=0.5\textwidth]{figures/scenario1.png}
%     \caption{Scenario 1. Scan unstructured documents and perform aggregation}
%     \label{fig:scenario1}
% \end{figure}

\hi{Scenario 2}: Analyzing Research Trends in NeurIPS (2023-2024). Cross-conference analysis of research trends between NeurIPS 2023 and 2024, involving retrieval, aggregation, and join operations.

% \begin{lstlisting}
% SELECT area, neurips_2024.count as 2024_count, neurips_2023.count as 2023_count
% FROM 
%   (SELECT PROMPT("Analyze the technical fields") as area, SEM_GROUP("Count the numbers of publications in each area") as count FROM FILE("neurips_2024.txt") GROUP BY area) neurips_2024 -- or use TABULAR
% LEFT JOIN
%   (SELECT PROMPT("Analyze the technical fields") as area, SEM_GROUP("Count the numbers of publications in each area") as count FROM FILE("neurips_2023.txt") GROUP BY area) neurips_2023 -- or use TABULAR
% ON SEM_MATCH(neurips_2024, neurips_2023, "the same technical area from both two documents", 0.9);
% \end{lstlisting}

\begin{figure}[htbp]
    \centering
    \includegraphics[width=0.50\textwidth]{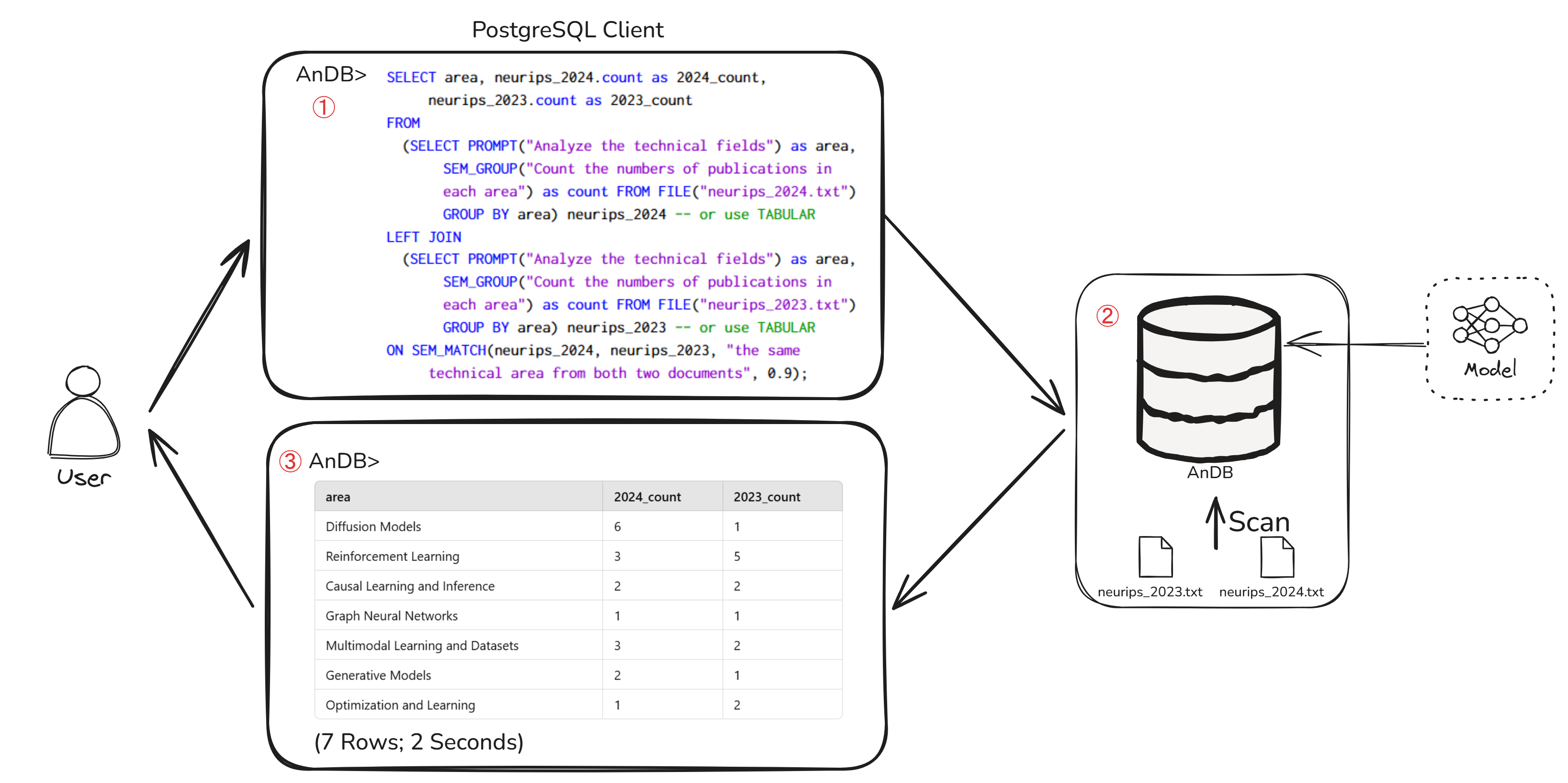}
    \caption{AnDB Workflow}
    \label{fig:workflow}
\end{figure}

Here, we demonstrate an end-to-end workflow for a use case as \autoref{fig:workflow} shown. \textbf{Step \textcircled{1}}: The user inputs an SQL-like query in the client interface. \textbf{Step \textcircled{2}}: AnDB scans the relevant unstructured data and invokes the LLM to perform the analysis. \textbf{Step \textcircled{3}}: The results are returned to the user, and presented in a structured format.

\begin{figure}[htbp]
    \centering
    \includegraphics[width=0.55\textwidth]{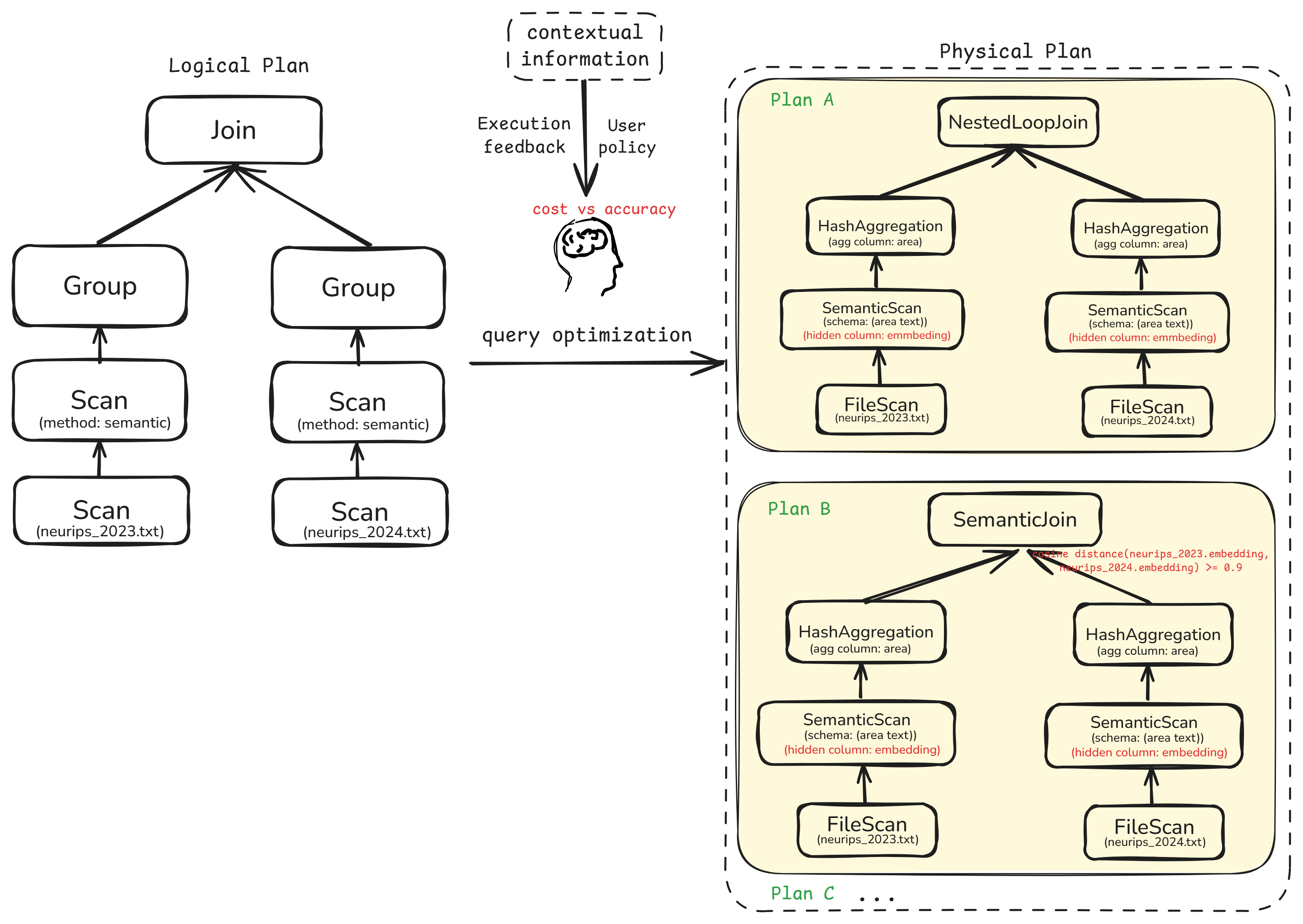}
    \caption{Scenario 2. Scan unstructured documents and perform aggregation and join}
    \label{fig:scenario2}
\end{figure}

The use case only involves two feasible plans in \autoref{fig:scenario2}. \textbf{Plan A}: First, unstructured data is transformed into tabular data, followed by aggregation using traditional join methods. \textbf{Plan B}: During the tabular transformation process, a domain-specific model is used to vectorize the corresponding text contents and embed them into a vector value in a hidden column. Subsequently, during the semantic join, vector similarity is utilized for judgment.

% \begin{acks}
% We would like to express our gratitude to David Anugraha and Sona Ghotra for providing us with early discussions and code contributions.
% \end{acks}

%\clearpage

\bibliographystyle{ACM-Reference-Format}
\bibliography{sample}

\end{document}